# Non-Hermitian exceptional physics in $\mathbb{R}P^2$ hyperbolic media

Shengyu Hu[1], Zhiwei Guo[1*], Wenwei Liu[2], Shuqi Chen[2,3,4], and Hong Chen[1]

[1.] MOE Key Laboratory of Advanced Micro-Structured Materials, School of Physics Science and Engineering, Tongji University, Shanghai 200092, China

[2.] The Key Laboratory of Weak Light Nonlinear Photonics, Ministry of Education, School of Physics and TEDA Institute of Applied Physics, Smart Sensing Interdisciplinary Science Center, Nankai University, Tianjin 300071, China

[3.] School of Materials Science and Engineering, Smart Sensing Interdisciplinary Science Center, Nankai University, Tianjin 300350, China

[4.] The Collaborative Innovation Center of Extreme Optics, Shanxi University, Taiyuan 030006 Shanxi, China

* Correspondence authors: 2014guozhiwei@tongji.edu.cn.

## Abstract

Conventional momentum space provides an orientable base space of a torus for topological classifications based on band theory. Here, we introduce a non-orientable momentum space isomorphic to the real projective plane $\mathbb{R}P^2$ within the low-symmetry media. We show that the local band fluidity can be characterized by an expanded dihedral group with non-Abelian properties, while the global band fluidity offers a versatile platform to explore the evolution of non-Hermitian exceptional manifolds, including order-1, higher-order, hybrid exceptional manifolds, diabolic points and even bound states in the continuum. Furthermore, the non-orientable momentum space can pave the way for exploring the emergence of phenomena for exceptional manifolds.

## Introduction

Ranging from the macro classic and micro quantum realm, symmetry serves as an elegant and intuitive paradigm of geometrization to depict the balance in the universe. A ubiquitous example is continuous space/time translation symmetry (STS/TTS), which anchors the conservation of momentum/energy [1]. In systems exhibiting these symmetries, isolated bands with real eigenvalues establish the initial image of trivial states based on topological band theory. When symmetry is appropriately reduced, band crossing or even band inversion occurs, indicating the topological phase transition towards nontrivial states [2, 3]. For STS, discrete versions in periodic media give rise to the concept of Brillouin zone (BZ) in momentum space, where boundaries with mirror symmetry define a base space isomorphic to a torus $\mathbb{T}^2$. Since the topology of base space has a profound inference over thereon bands, topological classifications are mainly executed on the orientable torus hitherto. Great success of topological band theory has spurred interest in non-orientable BZ, and recent focus is mainly laid on structures incorporating gauge field. For example, Möbius-twisted surface states can emerge when considering the synthetic dimensions associated with glide [4] or projective translation [5, 6] eigenvalues. Additionally, half of BZ can support a fundamental domain isomorphic to a Klein bottle [7]. However, the entire BZ remains inherently dominated by mirror symmetry. Worth mentioning, these pioneering works are mainly constructed within Hermitian systems with continuous TTS. For non-Hermitian systems [8–10], the breakdown of TTS originates from the non-equilibrium energy flux in the forms of either nonreciprocal coupling (channels between different portions in the system) or gain and/or loss (channels between the system and the environment), which leads to unique phenomena quite different from their Hermitian kinsfolks. A prominent example is exceptional manifolds (EMs), like exceptional points/lines/surfaces (EPs/ELs/ESs) [11–13], which can be regarded as spontaneous symmetry breaking due to unidirectional coupling between several identical portions of the system. Consequently, the equivalent eigenvalues (and eigenvectors) coalesce simultaneously, namely the system is detective.

Here, we study the intricate evolution of EMs in low-symmetry media that possesses a non-orientable momentum space with mirror and inversion symmetry simultaneously. On the $\mathbb{R}P^2$-like surface dominated by inversion symmetry, EMs carrying identical topological charges (TCs) merges, forming an order-1 EM. Conversely, on the $\mathbb{T}^2$-like surface dominated by mirror symmetry, EMs carrying opposite TCs merge, resulting in strings of bound states in the continuum (BICs)—zero EMs forms an anisotropic or hybrid EM, while nonzero EMs forms a diabolic or Dirac point. However, opposite TCs don't annihilation necessarily in the framework of revised band braiding theory. Interestingly, nonzero and zero EMs carrying identical topological charges can also merge, leading to the formation of an order-4 EM.

## Results

We begin with the interactions of light and media, which can be written as a linear-transformation form $H|\psi\rangle = \omega|\psi\rangle$ with

$$H = M^{-1}R = \begin{pmatrix} \overleftrightarrow{\varepsilon} & i\overleftrightarrow{\gamma} \\ -i\overleftrightarrow{\gamma} & \overleftrightarrow{\mu} \end{pmatrix}^{-1} \begin{pmatrix} 0 & i\nabla \times \\ -i\nabla \times & 0 \end{pmatrix}, \quad |\psi\rangle = \begin{pmatrix} \vec{E} \\ \vec{H} \end{pmatrix}. \quad (1)$$

Herein the curl matrix $R$ from the source-free Maxwell's equations describes the propagation of light, while the parameter matrix $M$ from the linear constitutive relation describes the response of media [14, 15]. A typical Hermitian formalism satisfied with $M^\dagger = M$ is nonmagnetic media without magneto-electric coupling, where permittivity $\vec{\varepsilon} = diag(a,b,c)$ (parameters $a, b, c \in \mathbb{R}$), permeability $\vec{\mu} = I_{3\times 3}$, and chirality $\vec{\gamma} = O_{3\times 3}$. A straightforward way to break Hermiticity is introducing several complex elements, either diagonal or off-diagonal. As an example, low-symmetry media $M_{NH} = \mathrm{M} + \Delta$ are detailed discussed, where all elements of $\Delta$ are zero except for $\Delta_{12} = \Delta_{21} = d$ ($d \in \mathbb{I}$), corresponding to monoclinic crystals [16]. The corresponding eigenvalues can be described as:

$$\Omega_{1,2} = 0, \Omega_{3,4,5,6} = \pm\sqrt{\frac{A \pm \sqrt{B}}{2c(ab-d^2)}}, \qquad (2)$$

where $A = \alpha_+ k_x^2 + \beta_+ k_y^2 + \gamma_+ k_z^2 + 2cdk_xk_y$ and $B = \alpha_-^2 k_x^4 + \beta_-^2 k_y^4 + \gamma_-^2 k_z^4 + 2\alpha_-\beta_- k_x^2 k_y^2 - 2\alpha_-\gamma_- k_x^2 k_z^2 + 2\beta_-\gamma_- k_y^2 k_z^2 + 4c^2 d^2 (k_x^2 + k_z^2)(k_y^2 + k_z^2) - 4cdk_xk_y[\alpha_- k_x^2 + \beta_- k_y^2 + (\alpha_- + \beta_-)k_z^2]$. Here $\alpha_\pm = ab \pm ac - d^2$, $\beta_\pm = ab \pm bc - d^2$, and $\gamma_\pm = c(a \pm b)$. Aside from the two zero eigenvalues $\Omega_{1,2}$ representing the static solutions [14], these nonzero eigenvalues are arranged as follows: $\Omega_3$: $(+,+)$, $\Omega_4$: $(+,-)$, $\Omega_5$: $(-,-)$, and $\Omega_6$: $(-,+)$, where the first (second) sign in the brackets denotes the choice of the first (second) sign in Eq. (2) outside (inside) the first square root [17].

For simplicity, we fix $a = 4, c = -3, d = 2i$ henceforth, since a single variable $b$ is sufficient to capture the intricate dynamics of EMs. In this setup, the eigenvalue spectra can be visualized as hypersurfaces homeomorphic to the composite base space $\mathbb{R}^4:\{\vec{k},b\}$. Additional constraints reduce the dimension of the base space, providing a local but intuitive prospective of the hypersurfaces. To start with, the constraint $b=4$ is taken to capture the system's symmetry in the momentum space. Considering the linear dependence $\Omega_3=-\Omega_5$ and $\Omega_4=-\Omega_6$, the iso-frequency contours (represented by the moduli of the real part) with twofold degeneracy are plotted in Fig. 1A, gradual red for $|Re(\Omega_3)|=|Re(\Omega_5)|$ and gradual blue for $|Re(\Omega_4)|=|Re(\Omega_6)|$ as a simplified projection of the hypersurfaces. The former modes, characterized by open contours, exhibit properties of hyperbolic materials. To depict the internal structure, we cut the hypersurfaces open with the longitudinal plane $k_y=0$ (red mesh) and the transverse plane $k_z=-1$ (blue mesh).

On the plane $k_y=0$, the Riemann sheets are continuous for both the real part (Fig. 1B) and the imaginary part (Fig. 1C) of each eigenvalue. This continuity is maintained by the mirror symmetry $\Omega(k_x)=\Omega(-k_x)$ and $\Omega(k_z)=\Omega(-k_z)$ from Eq. (2). Thereby, the base space here is isomorphic to a torus $\mathbb{T}^2$ like a conventional BZ. Noteworthy, there are two kinds of second-order eigenvalue coalescences embedded in the base space: (1) zero-energy coalescences with $A\pm\sqrt{B}=0$ for $\Omega_3=\Omega_5$ or $\Omega_4=\Omega_6$ (marked by green dashed lines); (2) nonzero-energy coalescences where only $B=0$ for $\Omega_3=\Omega_4$ and $\Omega_5=\Omega_6$ (marked by cyan dotted lines), which plays the role of fundamental elements in the evolutions of EMs described below. On the plane $k_z=-1$, these EMs are intersected and act as EPs. Intriguingly, the real parts of eigenvalues (Fig. 1D) remain continuous, while the imaginary parts (Fig. 1E) experience sign-flipping across the bulk Fermi arc [18]. This behavior can be attributed to the inversion symmetry $\Omega(k_x, k_y) = \Omega(-k_x, -k_y)$, which is composed of two sub-symmetry $\Omega(k_x) = \Omega^*(-k_x)$ and $\Omega(k_y) = \Omega^*(-k_y)$. This novel base space can be seen as a real projective plane $\mathbb{R}P^2$, as is shown in Fig. 1F. Akin to the Möbius strip [5, 6] and the Klein bottle [7], the real projective plane is also a non-orientable topological configuration. Despite the inevitable boundary self-intersections [19], the Morin surface can be perceived as an immersion of $\mathbb{R}P^2$ in a three-dimensional space (see Supplementary Materials, section 1.1, for details). Leveraging another auxiliary dimension, we can stretch the two-dimensional base space, twist 180° around the space center in this process, and get a spiral cylinder finally. In this configuration, each pair of antipodal points carrying equivalent eigenvalue spectra has been aligned with each other along the auxiliary dimension. We then bend the spiral cylinder and glue the upper and lower boundaries from inside (Supplementary Movie S1). The surface of the spiral cylinder forms Morin surface, corresponding to the base space shown in Fig. 1 D–F. Compared to a torus, this base space endows spectra with unique topological properties in the dynamics of EMs. Take the spectrum vorticity of EPs as an example, which is described by the winding number $v_{mn}(C) = -\frac{1}{2\pi} \oint_C \nabla_{\vec{k}} \arg[\Omega_m(\vec{k}) - \Omega_n(\vec{k})] \cdot d\vec{k}$ [20]. To ensure this vorticity invariable when an EP moves in the base space, we define a closed loop C encircling the EP and the normal vector of the base space [away from the Γ point: $\vec{k}_\Gamma = (0,0,0)$] meet with the right-hand rule [21]. By tracing different clockwise closed loops C on the plane $k_z = -1$, a total accumulated phase of $\pm\pi$ can be obtained, indicating half-integer topological charges $v_{43} = \pm 1/2$ for nonzero EPs (Fig. 1G) and $v_{35} = \pm 1/2$ for zero EPs (Fig. 1H). We notice the inversion symmetry ensures a pair of EPs with identical vorticities, positive along the axis $k_x$ and negative along the axis $k_y$. In stark contrast, EPs and their mirror partners always carry opposite vorticities, e.g. on a torus like Fig. 4. Besides, there exists phase jumping across the high-symmetry lines connecting identical topological charges, marking the bulk Fermi arcs. (Moreover, the degeneracy of imaginary parts between eigenvalues is defined as the imaginary bulk Fermi arc [22].)

Next, we restore the fluidity of the hypersurfaces along the axis of parameter $b$, and demonstrate the dynamics of EMs in the momentum space (Supplementary Movie S2). Viewed as a slowly-varying time indicator, parameter $b$ ranges from 50 (an approximation to infinite) to -50. The local fluidity with fixed momentum $\vec{k}_0$ is manifested as band braiding (Fig. 2C), which can be described by the transition $\begin{bmatrix} \Omega_3(b=50) \\ \Omega_4(b=50) \end{bmatrix} = T \cdot \begin{bmatrix} \Omega_3(b=-50) \\ \Omega_4(b=-50) \end{bmatrix}$. Based on the conditions $\{\Omega(b \to +\infty)\} = \{\Omega(b \to -\infty)\} \in \mathbb{R} \cup \mathbb{I}$ from Eq. (2), the nonzero eigenvalues from the initial and final spectra are restricted on two orbits, and each orbit contains four permitted energy levels on the complex plane $\text{Re}(\Omega) - \text{Im}(\Omega)$ (Fig. 2B). (Without loss of generality, we choose two linearly independent eigenvalues $\Omega_3$ and $\Omega_4$ to make up the basis vector of transition.) Hence, the transition can be decomposed into operations of interchanging between two orbits and/or local rotating between four energy levels on one orbit. And the transfer matrix $T$ linking the initial and final spectra belongs to an expanded dihedral group $\mathbb{D}_4 : \{C_m\} = \{c_m\} \otimes \{\pm 1, \pm i\}$ described by $C_m = (c_m, \varsigma)$. Wherein quasi-group $\{c_m\}$ comprises eight elements:

$$c_1 = \begin{pmatrix} 1 & 0 \\ 0 & 1 \end{pmatrix}, c_2 = \begin{pmatrix} 0 & 1 \\ 1 & 0 \end{pmatrix}, c_3 = \begin{pmatrix} 0 & 1 \\ -i & 0 \end{pmatrix}, c_4 = \begin{pmatrix} 1 & 0 \\ 0 & -i \end{pmatrix},$$

$$c_5 = \begin{pmatrix} 1 & 0 \\ 0 & -1 \end{pmatrix}, c_6 = \begin{pmatrix} 0 & 1 \\ -1 & 0 \end{pmatrix}, c_7 = \begin{pmatrix} 0 & 1 \\ i & 0 \end{pmatrix}, c_8 = \begin{pmatrix} 1 & 0 \\ 0 & i \end{pmatrix}. \tag{3}$$

Here, the quasi-group, a generalized algebraic structure newly-defined from group, satisfies the conditions with one identity element ($c_1$) and associative law, but without inverse elements and closure. Meanwhile, the group $\{\pm 1, \pm i\}$ corresponds to the global rotation $\varsigma$ of the entire spectra. When this global rotation can be eliminated or neglected, e.g. take ratio $\Omega_4/\Omega_3$ as the normalized factor or take eigenstate $|\psi\rangle$ in Eq. 1 as the basis vector, group $\{C_m\}$ with elements $C_m \propto c_m$ can restore the closure and retrieve the prototypical fourth-order dihedral group. Intriguingly, the above-mentioned operations don't commute with each other, leading to the non-Abelian of group $\mathbb{D}_4$ and $\{c_m\}$ evidentially shown in the Cayley table (Table S2 in the Supplementary Materials). This non-Abelian can be illustrated by a spiral stair model (e.g. $C_2 \cdot C_4 = C_8 \cdot C_2 \neq C_4 \cdot C_2$): Starting from the middle of the stair (the darkest region in Fig. 2B), one goes upward counterclockwise while another goes downward clockwise (towards the brighter region). Their positions are associated through the operation of interchanging $C_2$ (the starting point $C_2 \cdot C_1 = C_2$ and the terminal point $C_2 \cdot C_4 = C_7$), but their directions are always opposite (the upward one $C_4 \cdot C_1 = C_4$ and the downward one $C_8 \cdot C_2 = C_7$), corresponding to the operations of local rotation with conjugated elements $C_4$ and $C_8$. In previous works, three or more bands are necessary in the Hamiltonian to realize non-Abelian topology, such as quaternion group $\mathbb{Q}_8$ in Hermitian systems [23–25] and braiding group $\mathbb{B}_N$ ($N \geq 3$) in non-Hermitian systems [26–28]. However, the dihedral group $\mathbb{D}_4$ may provide a heuristic perspective of non-Abelian in systems with just two bands. (One intuitive example is the subsystem only with eigenvalues of $\Omega_{3,4}$.) In fact, when tracing an open path in the base space, unoccupied but permitted bands can also participate in the process of band braiding, which may reduce the system complexity for applications like logic gates [29]. A complete set logic gates from group $\{C_m\}$ can be selected near the high-symmetry lines (Fig. 2C). Noteworthy, the logic gates are ill-defined occasionally by the fidelity, since the transfer matrix $T$ is: (1) multivalued near point P due to merging of two orbits, and nonzero-energy coalescence occurs at $b = -1$ [Fig. 2, C1 and C5]; (2) invalid near point O due to vanishing of one orbit, and zero-energy coalescence occurs at $b = \pm\infty$ (see Supplementary Materials, section 3.1, for details) of the constitutive matrix in Eq. 1. In addition, other EPs seem to be indistinguishable [marked by diamonds in Fig. 2, C2–C3 and C6–C7]. It is that the local spectra cannot reveal the properties of EPs completely. We will discuss their detail classification aided by the global spectra in the next section.

On the other hand, the global fluidity is constrained by the Hamiltonian self-similarity $H(\eta\vec{k}, b) = \eta H(\vec{k}, b)$ in Eq. (1): When altering $b$, the travelling EMs, composed of EPs with the same eigenvectors, maintain stable configurations of straight lines originating from point Γ. In this context, we can employ a mathematical treatment named box quantization [30]. As a natural promotion of Fig. 1, the evolutionary trajectory of EMs can be completely captured by the intersections (i.e., EPs) on the surface of a cube domain $\vec{k} \in [-1,1]^3$. This domain can be further reduced to $[0,1]^3$ due to above-mentioned symmetry, as is shown in Fig. 2A. On its top surface, the journey begins at point O: a pair of nonzero EPs $\chi_{1\pm}^{NZ}$ carrying opposite topological charges is created here at the initial moment $b \to +\infty$. With the reduction of $b$, both EPs move along the axis of $k_x$ towards point Z, the center of inversion symmetry, in Fig. 3A. Here, the faster $-1/2$ charge $\chi_{1-}^{NZ}$ (with trajectory marked by black dashed lines) pioneers collision with its inversion partner $\chi_{2-}^{NZ}$ coming from point O′ when $b = 8$. Due to the conservation law, this collision generates the first kind of merging EMs, which possess integer spectrum vorticity $-1$. Hitherto, the above-discussed topological properties have primarily focused on eigenvalues. In fact, similar merging phenomenon can also be observed in the view of eigenvectors, through the polarization vorticity $q(C) = \frac{1}{2\pi} \oint_C \frac{\nabla_{\vec{k}} \arg[S_1(\vec{k}) + iS_2(\vec{k})]}{2} \cdot d\vec{k}$ [31]. Herein $S_i$ is the $i$-th Stokes parameter, and $\frac{\arg[S_1(\vec{k}) + iS_2(\vec{k})]}{2}$ describes the orientation angle of the major axis of the polarization ellipse. At $b = 8$, point $Z$ corresponds to a left-handed circularly polarized point ($C$ point) with an integer polarization vorticity 1 (Fig. 3B). To illustrate the asymptotic dispersion relation around point $Z$, we assume a small and real scalar value for $\delta k$ as a disturbance [32], and the response at the direction $\vec{k}$ can be described as $\Omega_i\left(\vec{k}_Z + \delta k \cdot \frac{\vec{k}}{|\vec{k}|}\right) - \Omega_i(\vec{k}_Z) = Re(\delta \Omega_i) + i \cdot Im(\delta \Omega_i) \propto \delta k^{\epsilon_{Re}} + i \cdot \delta k^{\epsilon_{Im}}$ (see Supplementary Materials, section 3.3, for details). We notice the linear dispersion exists in all directions, except: (1) $\vec{k} = [\pm 1,0,0]$, the bulk Fermi arc has quadratic dispersion with $\epsilon_{Re} = 2$ and $\epsilon_{Im} = 1$; (2) $k_x = 0$, the imaginary bulk Fermi arc is dispersionless with $\epsilon_{Re} = 1$ and $\epsilon_{Im} = 0$ (Fig. 3E). In that case, it corresponds to an order-1 EM. Similar situations appear for the slower $+1/2$ charge $\chi_{1+}^{NZ}$ and its inversion partner $\chi_{2+}^{NZ}$ (with trajectory marked by white solid lines), which forms an order-1 EM (marked by the cyan diamond) at $b = 0$, corresponding to a right-handed $C$ point with +1 polarization or spectrum vorticity (Fig. 3A). However, this order-1 EM is not as stable as its components—two EPs ($\chi_{1+}^{NZ}$ and $\chi_{2+}^{NZ}$ or $\chi_{1-}^{NZ}$ and $\chi_{2-}^{NZ}$) parted with each other subsequently, and move along the axis of $k_y$ away from point Z, cross the boundary of the top surface, and enter the front surface of the cube domain in Fig. 2A. Intriguingly, the symmetry of mirror, rather than inversion, plays the primary role on the front surface. Hence, the inversion partner $\chi_{2-}^{NZ}$ ($\chi_{2+}^{NZ}$) of the charge $\chi_{1-}^{NZ}$ ($\chi_{1+}^{NZ}$) converts to $\chi_{2+}^{NZ}$ ($\chi_{2-}^{NZ}$) with opposite charge, marked by crosses in Figs. 3A and 4A1. Such charge conversion can be seen as the result crossing the Möbius-like boundary (red arrows) in the $\mathbb{R}P^2$ base space.

Focusing on the zero EP $\chi_{1-}^{Z}$ (with trajectory marked by red triangles in Fig. 4A1) moving along axis of $k_z$ towards point $Y$, the nonzero EP $\chi_{1-}^{NZ}$ (with trajectory marked by black dashed lines) catches up and collides with it at point $P$ when $b = 1$. It engenders an order-4 EM (marked by the spring green hexagram) with quarter dispersion (Fig. 4E). Before this collision, we notice the half integer vorticity of the nonzero EP $\chi_{1-}^{NZ}$ is allocated in $v_{43}$ (Fig. 4A1) and $v_{45}$ (Fig. 4A2) (see Supplementary Materials, sections 3.3, 4.1 and 4.2, for details). After this collision, the zero-energy coalescence $\chi_{1-}^{Z}$ between from $\Omega_3$ and $\Omega_5$ [like Fig. 1, D–E] is transferred to $\Omega_4$ and $\Omega_6$ [like Fig. 4, F–G]. And the renewed zero EP $\chi_{1-}^{Z}$ carries on its journey towards point Y, the center of mirror symmetry. At point Y, it meets its mirror partner $\chi_{2+}^{Z}$ (with trajectory marked by blue triangles) when $b = 0$. Focusing on the dispersion of the imaginary part, it has a slope of 1/2 like ordinary order-2 EP, except on the directions $k_x = 0$ with linear dispersion (Fig. 4I), corresponding to the hybrid EM [20] or the anisotropic EM [33]. The spectrum (Fig. 4A3) vorticity is eliminated due to the annihilation of opposite charges carried by $\chi_{1-}^{Z}$ and $\chi_{2+}^{Z}$ with properties of bound state in the continuum (BIC) [34]. However, similar annihilation does not always happen for opposite charges—an archetypal example is the mirror pair of nonzero EPs $\chi_{1-}^{NZ}$ and $\chi_{2+}^{NZ}$. When $b = -4/7$, their collision spawns a novel vortex texture around the ill-defined point Y with all-zero Stokes parameters (Fig. 4L), corresponding to a V point. For eigenvectors, their orthogonality is restored at point Y, corresponding to a diabolic point [17] or a Dirac point [20], and the Hamiltonian is non-defective now. Take $\Omega_3 = \Omega_5 = -0.5774i$ at point Y as an example, their eigenvectors are $[-0.2481i, 0.8682, 0, 0, 0, -0.4297]^T$, a propagation mode, and $[0, 0, -0.5i, 0.8660, 0, 0]^T$, a bound mode, which can come back to the classic model of guided resonance and BIC at the high symmetry point due to coupling. As a comparison, both of two eigenvectors when $b = 0$ are $[0, 1, 0, 0, 0, 0]^T$, a coalesced bound mode. Worth mentioning, the pure imaginary eigenvalues seem to be elusive. In fact, it corresponds to be pure real for evanescent waves with imaginary momentum due to the self-similarity. On the other hand, the self-similarity also extends the BIC as a line in the momentum space. For eigenvalues, linear dispersion exists around the diabolic point, except on the directions $k_x = 0$ with quadratic dispersion, which may come from the parabola on the longitudinal section of the Dirac cone (Fig. 4M). Intriguing, the theory of eigenvalue braid used to analysis analogous non-annihilations [35] becomes invalid in this novel situation. In Fig. 1E, we notice the imaginary parts experience sign-flipping across the bulk Fermi arc, which break the requirements of continuity of braiding elements [36]. For a single EP, the bulk Fermi arc is unilateral, and there exists an enclosed path avoiding crossing it directly. A common strategy is to select the point on the bulk Fermi arc as the starting and terminal points [37]. For a merging EP composed multiple single EPs, the bulk Fermi arc tends to be bilateral and unavoidable through path selection. Nevertheless, we can leverage the mapping $\{\Omega_3, \Omega_4, \Omega_5, \Omega_6\} \rightarrow \{\Omega_3^2, \Omega_4^2, -\Omega_5^2, -\Omega_6^2\}$ to ensure the continuity of braiding to some extent (see Supplementary Materials, section 3. 2, for details). Then, $\chi_{1+}^{\text{NZ}}$ and $\chi_{2+}^{\text{NZ}}$ collide at the pole point $P$ when $b = -1$. The eigenvalues are divergent [38], since the parameter matrix $M$ is singular ($ab - d^2 = 0$) in Eq. 1. After that, EMs break away the high symmetry lines, keep on their journey on the surface, and come back to the original point O.

**Discussion**

We have observed emergent phenomena from the evolution of non-Hermitian bands in the monoclinic crystals, where EMs generate, merge, and annihilate in the non-orientable Morin surface. Similar inversion symmetry has been observed in beta-phase $Ga_2O_3$, and triggered hyperbolic shear polaritons [16]. As a stark contrast, mirror symmetry is dominate in chiral media thoroughly (see Supplementary Materials, section 2.2, for details). Take thriving applications of exceptional physics into account, such as sensing [32], lasing [39], mode switching [40], and hardware encryption [41], $\mathbb{R}P^2$ hyperbolic media are expected to develop into a powerful platform convenient to fabricate, tiny in footprint, flexible on usage and diverse for functions.

## Data availability

All the data that support the findings of this study are available from the corresponding authors upon reasonable request.

## Code availability

All the codes that support the findings of this study are available from the corresponding authors upon reasonable request.

## Acknowledgments

This work is partially supported by the National Key Research Program of China (Nos. 2021YFA1400602, and 2023YFA1407600), the National Natural Science Foundation of China (Grant Nos. 12004284, 12274325, and 12374294), and the Shanghai Chenguang Plan (Grant No. 21CGA22).

## Author contributions

Z. Guo conceived the idea. S. Hu proposed the model, performed the numerical simulations and theoretical analyses. Z. Guo, S. Chen, and H. Chen supervised the whole project. S. Hu, Z. Guo, and W. Liu wrote the manuscript. All authors contributed to discussions of the results and the manuscript.

## Competing interests

The authors declare no competing financial and non-financial interests.

## Additional information

Supplementary materials to this article can be found online at ***.

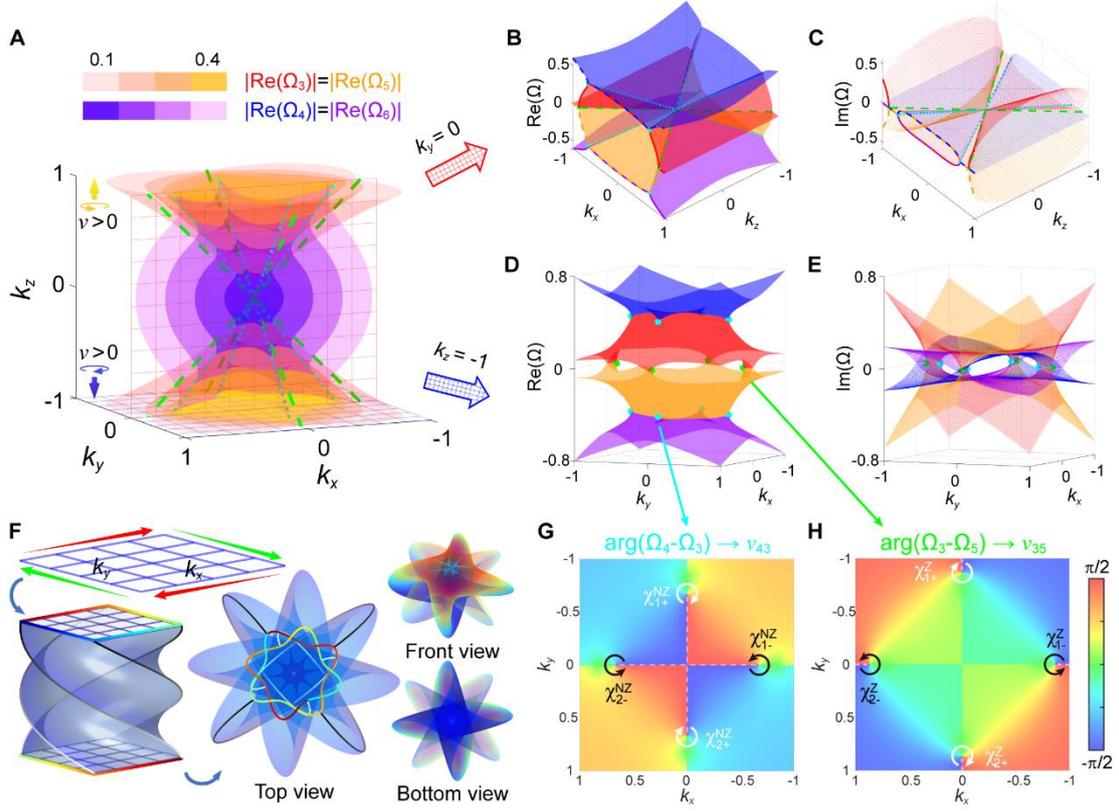

Figure 1. **Band symmetry of $\mathbb{R}P^2$ hyperbolic media.** (**A**) Iso-frequency contours with twofold degeneracy when $b=4$, gradual red for $|Re(\Omega_3)| = |Re(\Omega_5)|$ and gradual blue for $|Re(\Omega_4)| = |Re(\Omega_6)|$. Red and blue meshed planes denote $k_y = 0$ with the mirror symmetry and $k_z = -1$ with the inversion symmetry, respectively. (**B** and **C**) Real (B) and imaginary (C) parts of the eigenvalues on the plane $k_y = 0$, embedding four zero ELs (green dashed lines) and four nonzero ELs (cyan dotted lines). (**D** and **E**) Real (D) and imaginary (E) parts of the eigenvalues on the plane $k_z = -1$, embedding four zero EPs (green points) and eight nonzero EPs (cyan points). (**F**) $k_z = -1$ with the inversion symmetry. The boundaries with the same color should be identified along the marked direction, which is essentially the Morin surface after stretching, twisting, bending, and gluing. (**G** and **H**) Spectrum vorticities $v_{43}$ for the nonzero EPs (G) and $v_{35}$ for the zero EPs (H). On the surface $k_z = -1$,

clockwise (anticlockwise) for positive (negative) topological charges, as the blue arrow in (A). White dashed lines are the bulk Fermi arcs.

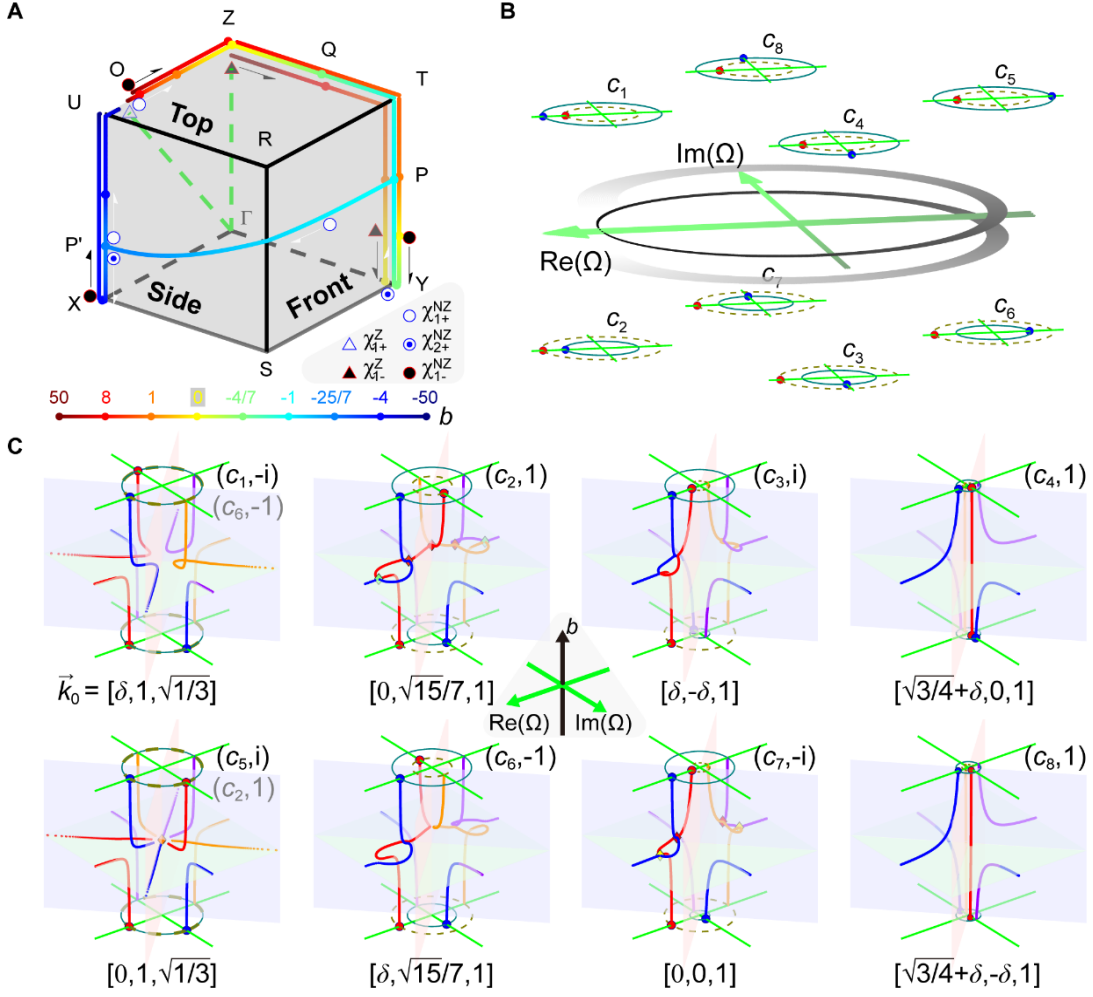

Figure 2. **Non-Abelian local fluidity described by dihedral group** $\mathbb{D}_4$. (**A**) Evolution trajectories of EMs dependent on $b$ on the surface of a cube domain $\vec{k} \in [0,1]^3$, where trajectories of nonzero EPs, $\chi_{1+}^{NZ}$ (white circle), $\chi_{1-}^{NZ}$ (black circle) and $\chi_{2+}^{NZ}$ (white dotted circle), are plotted outside the surface, and trajectories of zero EPs, $\chi_{1+}^{Z}$ (white triangle) and $\chi_{1-}^{Z}$ (black triangle), are plotted inside the surface for clarity. (**B**) Eigenvalues $\Omega_3(b \to \pm\infty)$ as red points and $\Omega_4(b \to \pm\infty)$ as blue points are restricted on two orbits, and each orbit has four permitted energy levels on the complex plane $Re(\Omega) - Im(\Omega)$. Dihedral group $\mathbb{D}_4$ illustrated by a spiral stair model. On the stair, steps with the same color are linked with the operation of orbit interchanging, like $c_4$ and $c_7$, while adjacent steps are linked with the operation of local rotating, like $c_1$ and $c_4$. (**C**) Local fluidities near point P (c1 and c5), point

Q (c2 and c6), point Z (c3 and c7), and point O (c4 and c8) in (A), where $\delta = 0.01$. $\mathbb{D}_4$ can depict the transformation between the initial and final spectra, and orbit merging leads to multivalued transformations in (c1 and c5). EPs are marked by diamonds with colors of related $b$ in (A).

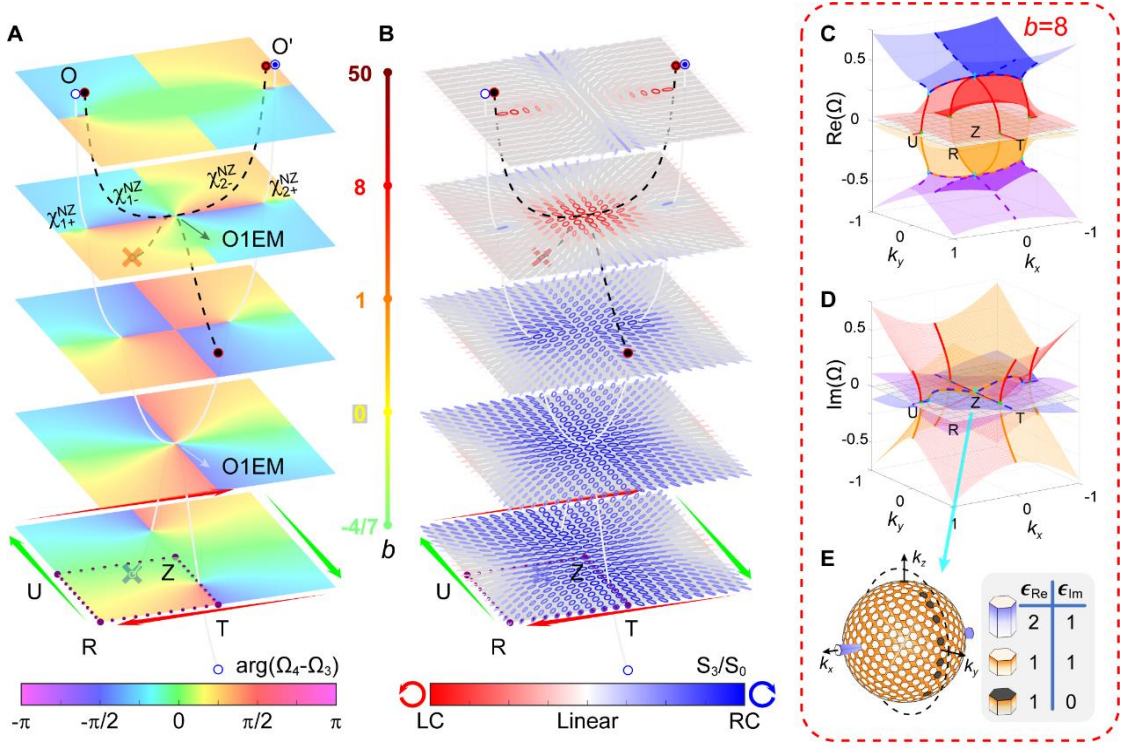

Figure 3. **Global fluidity dominated by inversion symmetry on the top surface $k_z = 1$.**
(**A** and **B**) Evolutions of spectrum (A) and polarization (B) vorticities dependent on $b$. White solid and black dashed lines denote trajectories of nonzero EPs carrying positive topological charges, $\chi_{1+}^{NZ}$ (white circle) or $\chi_{2+}^{NZ}$ (white dotted circle), and negative topological charges, $\chi_{1-}^{NZ}$ (black circle) or $\chi_{2-}^{NZ}$ (black dotted circle), respectively. O1EM, Order-1 EM. (**C** to **E**) When $b=8$, real (C) and imaginary (D) parts of the eigenvalues, which are highlighted for clarity in its quarter region $[k_x, k_y] \epsilon [-1,0]^2$ and at the high symmetry lines $k_{x,y} = 0$. Merging of inversion partners $\chi_{1-}^{NZ}$ and $\chi_{2-}^{NZ}$ forms an order-1 EM (cyan diamonds) at point Z with linear dispersions of the eigenvalues (E). For the Fibonacci cluster diagram composed of hexagonal prisms, the length and side color of prisms indicate the dispersion $\epsilon_{Re}$ of the real part, the auxiliary line and facet color of prisms indicate the dispersion $\epsilon_{Im}$ of the imaginary part, and the direction of prisms indicates the direction of disturbance $\delta k$. At $\vec{k} = [\pm 1, 0, 0]$, the bulk

Fermi arc has quadratic dispersion with $\epsilon_{Re} = 2$; at $k_x = 0$, the imaginary bulk Fermi arc is dispersionless with $\epsilon_{Im} = 0$.

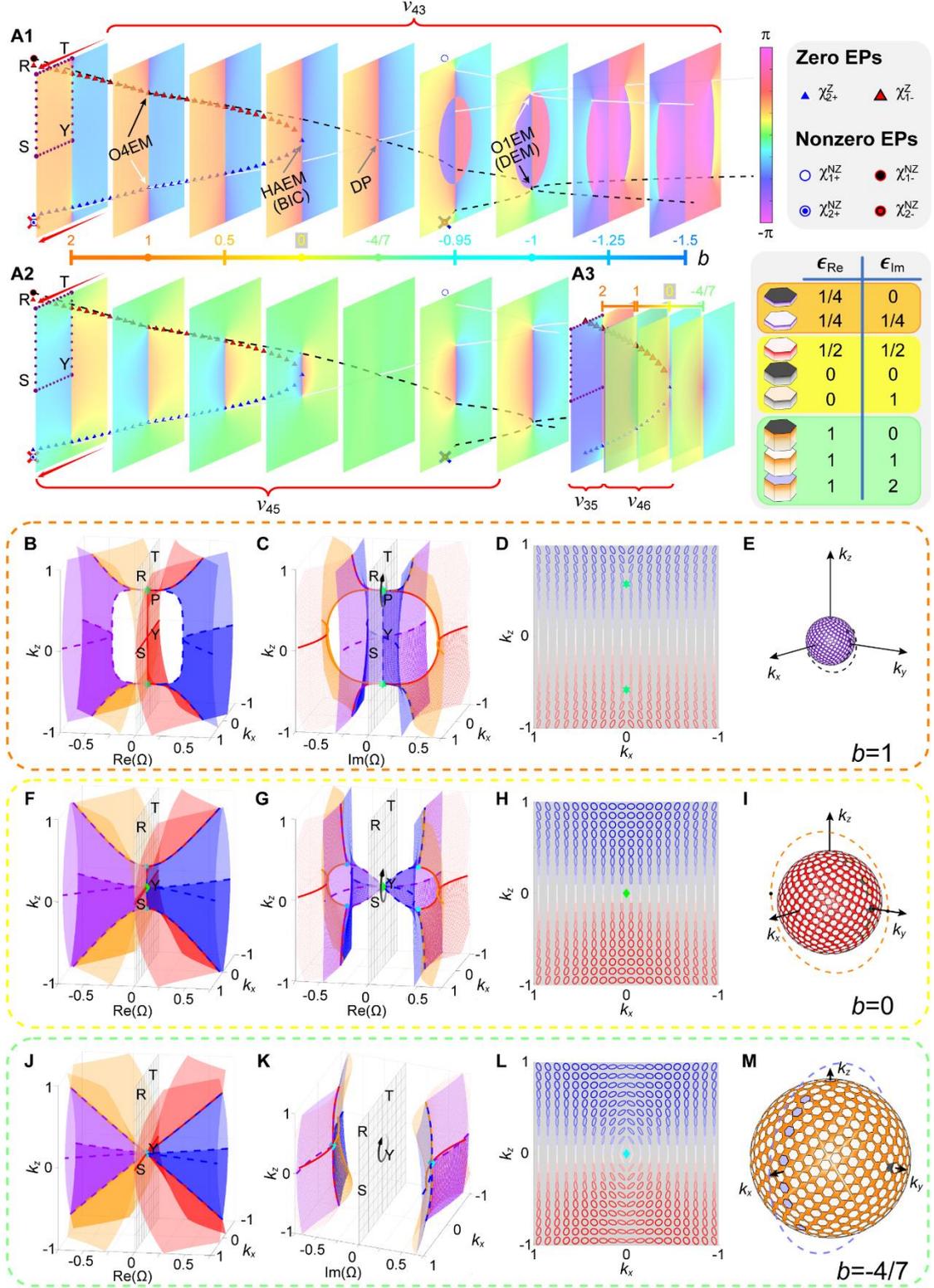

Figure 4. **Global fluidity dominated by mirror symmetry on the front surface** $k_y = 1$. (**A**) Evolutions of spectrum vorticities dependent on $b$. Blue and red triangles denote trajectories of zero EPs, $\chi^Z_{2+}$ with positive topological charges and $\chi^Z_{1-}$ with negative topological charges, respectively. The red dashed boundaries in Fig. 3(A) and Fig. 4(A)

identified along opposite directions. Crossing this Möbius-like boundary, $\chi_{2-}^{NZ}$ ($\chi_{2+}^{NZ}$) converts to $\chi_{2+}^{NZ}$ ($\chi_{2-}^{NZ}$) with opposite charge, which is marked by the red (blue) cross. O4EM, Order-4 EM; HAEM, hybrid or anisotropic EM; DP, diabolic or Dirac point; DEM, divergent EM. (**B** to **E**) When *b*=1, real (B) and imaginary (C) parts of the eigenvalues, which are highlighted for clarity in its quarter region $[k_x, k_z]\epsilon[-1,0]^2$ and at the high symmetry lines $k_{x,z} = 0$. Merging of $\chi_{1-}^{NZ}$ and $\chi_{1-}^{Z}$ forms an order-4 EM (spring green hexagrams) at point P with quarter dispersions of the eigenvalues (E). Encircling the EP along the path $\vec{k} = \{sin\theta, 1, \sqrt{1/3} + cos\theta\}$, polarization vorticity is 0.25, (D). (**F** to **I**) When *b*=0, merging of mirror partners $\chi_{1-}^{Z}$ and $\chi_{2+}^{Z}$ forms a hybrid EM (green diamonds) at point Y, which has a slope of 1/2 like ordinary order-2 EP, except on the directions $k_x = 0$ with linear dispersion (I). Encircling the EP along the path $\vec{k} = \{sin\theta, 1, cos\theta\}$, polarization vorticity is zero, corresponding to a BIC (H). (**J** to **M**) When *b*=-4/7, merging of mirror partners $\chi_{1-}^{NZ}$ and $\chi_{2+}^{NZ}$ forms an DP (cyan diamonds) at point Y with linear dispersion of the eigenvalues (M). Encircling the EP along the path $\vec{k} = \{sin\theta, 1, cos\theta\}$, polarization vorticity is 0.5 (L).